\documentclass[twocolumn,showpacs,prb,aps,amsmath,amssymb,superscriptaddress]{revtex4}

\usepackage[dvipdfmx]{graphicx}% Include figure files
\usepackage{bm}% bold math
\usepackage[dvipdfmx]{color}

\usepackage{txfonts}

\begin{document}

\title{
Tree-Graph Based Construction of Quantum Spin Models with Exact Ground State
}

\author{Toshiya Hikihara$^*$}
\affiliation{Graduate School of Science and Technology, Gunma University, Kiryu, Gunma 376-8515, Japan}

\begin{abstract}
We propose a protocol to generate an antiferromagnetic $S=1/2$ Heisenberg model with the exact ground state based on a tree graph.
The generated model has a correspondence with a tree graph and possesses the product state of singlet dimers as its unique ground state.
A procedure for constructing a model with exact, massively degenerate ground states is also introduced.
\end{abstract}

%%% Keywords are not needed any longer. %%%
%%%\kword{keyword1, keyword2, keyword3, \ldots}
%%%

\maketitle

\section{Introduction}

Exact results are particularly valuable in the study of quantum many-body physics.
They can serve to verify analytical predictions and also provide a starting point for approximation theories.
The most famous example of exact results for the quantum many-body problem is the Majundar-Ghosh model\cite{MajumdarG1969,Majumdar1970} for the spin-1/2 system on a zigzag ladder lattice.
It has been shown that the Majundar-Ghosh model is expressed as a sum of projection operators and has dimer product states as its doubly degenerate ground states.
%The Affleck-Kennedy-Lieb-Tasaki model\cite{AffleckKLT1987} is another famous example for which the exact ground state is obtained.
Other famous models where the exact ground-state solution has been established include the Affleck-Kennedy-Lieb-Tasaki model\cite{AffleckKLT1987}, the Shastry-Sutherland model\cite{ShastryS1981,VerkholyakSMS2014}, and the Richardson model\cite{DukelskyPS2004,DelftR2001}.
Additionally, various quantum spin systems with unique, finitely-degenerate, or infinitely-degenerate exact ground states have been proposed.\cite{NakaoT1995,NakaoT1997,TakanoKS1996,VerkholyakS2013,MoritaS2016,BoseG1997,SchmidtL2010,Xian1995,VerkholyakS2012,HikiharaTOS2017,Takano1994,Kumar2002}

In this work, we propose a protocol for systematically generating quantum spin models with exact ground states based on tree graphs.
The model considered is the $S=1/2$ antiferromagnetic Heisenberg model, whose Hamiltonian is given by
\begin{eqnarray}
\mathcal{H} = \sum_{i,j} J_{ij} {\bm s}_i \cdot {\bm s}_j,
\label{eq:Ham}
\end{eqnarray}
where ${\bm s}_i$ is the $S=1/2$ spin operator at the $i$th site.
We suppose that the number of spins in the model, $N$, is even.
The exchange interaction constants $J_{ij}$ are assumed to be positive, i.e., $J_{ij} > 0$.
The spatial configuration of $\{ J_{ij}\}$ which has a one-to-one correspondence with a given tree graph is determined by the protocol we propose.
We then show that when the exchange interaction constants $\{ J_{ij} \}$ satisfy a specific condition, the model (\ref{eq:Ham}) possesses a singlet-dimer product state as its unique ground state.
We also demonstrate that by relaxing the condition on $\{ J_{ij}\}$, one can construct a model with the macroscopically degenerate ground states.

The paper is organized as follows.
In Sec.\ \ref{sec:model}, we introduce the protocol for generating the quantum spin model from a given tree graph.
We then show in Sec.\ \ref{sec:ExactGS} that the generated model has a singlet-dimer product state as the unique ground state when the exchange interaction constants satisfy a specific condition.
The case where the model has massively degenerate ground states is also discussed.
Section\ \ref{sec:summary} is devoted to a summary and conclusion.

\section{Model and Tree Graph}\label{sec:model}

We consider the spin-1/2 Heisenberg model (\ref{eq:Ham}) that has a correspondence with a tree graph.
The exchange interaction constants $\{ J_{ij} \}$ are generated according to a tree graph as follows.

\begin{figure}
\includegraphics[width=60mm]{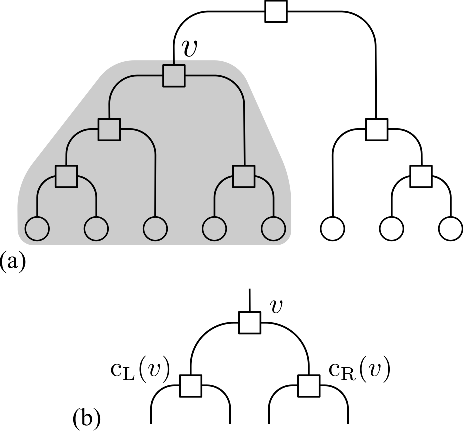}
\caption{
(a) Example of a generic binary tree graph for an eight-spin system.
Squares and circles denote vertices and bare $S=1/2$ spins, respectively.
The grey area represents the spin block assigned to the vertex $v$ indicated in the panel.
(b) A vertex $v$ and its child vertices $c_{\rm L}(v)$ and $c_{\rm R}(v)$.
The block spins $\tilde{S}_{c_{\rm L}(v)}$ and $\tilde{S}_{c_{\rm R}(v)}$ are coupled via an exchange interaction with the exchange constant $2J_v$.
}
\label{fig:treegraph}
\end{figure}

Figure\ \ref{fig:treegraph}(a) presents an example of a generic tree graph.
In this study, we focus on binary trees; that is, internal vertices (vertices other than leaves) have two child vertices and one parent vertex, except the root vertex, which has only two child vertices.
At the boundary of the tree graph, there are leaves, where the $S=1/2$ spins are located.

Each internal vertex has a corresponding spin block composed of the spins on its descendant leaves.
[See Fig.\ \ref{fig:treegraph}(a).]
We express the block spin operator of the spin block corresponding to the vertex $v$ as
\begin{eqnarray}
\tilde{\bm S}_v = \sum_{i \in v} {\bm s}_i,
\label{eq:blockspin}
\end{eqnarray}
where $\sum_{i \in v}$ is taken for the spins in the spin block assigned to the vertex $v$.

\begin{figure*}
\includegraphics[width=130mm]{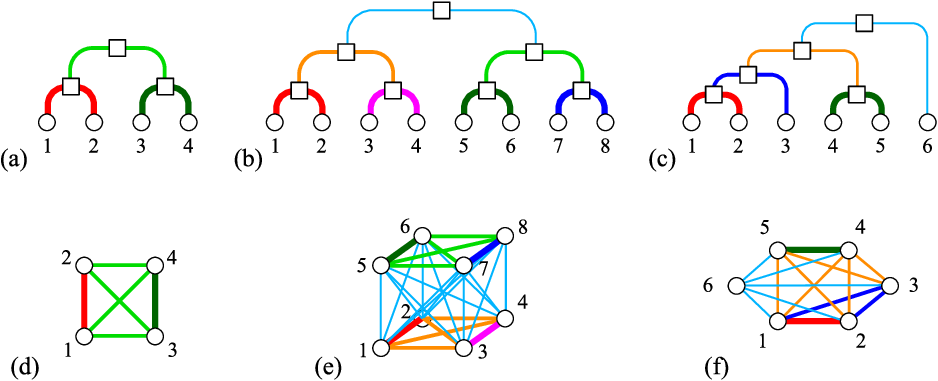}
\caption{
(Color online) 
Correspondence between tree graphs and quantum spin models.
(a) Perfect-Binary Tree (PBT) with two layers, (b) PBT with three layers, and (c) a generic tree graph.
In (a), (b), and (c), squares and circles represent the vertices and bare spins, respectively.
(d), (e), and (f) depict the quantum spin models corresponding to (a), (b), and (c), respectively.
In (d), (e), and (f), circles and lines denote the $S=1/2$ spins and exchange interactions, respectively.
}
\label{fig:tree_and_model}
\end{figure*}

To construct the spin model, we allocate a positive exchange interaction constant $2 J_v > 0$ to each internal vertex $v$.
We then include in the spin model the exchange interaction between the block spin operators of two child vertices
\begin{eqnarray}
h_v = 2 J_v \tilde{\bm S}_{c_{\rm L}(v)} \cdot \tilde{\bm S}_{c_{\rm R}(v)},
\label{eq:int_vertex}
\end{eqnarray}
where $c_{\rm L}(v)$ and $c_{\rm R}(v)$ represent the two child vertices of the vertex $v$ [see Fig.\ \ref{fig:treegraph}(b)].
The spin model we consider is given by the sum of such exchange interaction $h_v$ over all internal vertices as
\begin{eqnarray}
\mathcal{H} &=& \sum_v 2 J_v \tilde{\bm S}_{c_{\rm L}(v)} \cdot \tilde{\bm S}_{c_{\rm R}(v)}
\nonumber \\
&=& \sum_v 2 J_v \sum_{i \in c_{\rm L}(v)}\sum_{j \in c_{\rm R}(v)}
{\bf s}_i \cdot {\bf s}_j.
\label{eq:Ham_totalspin}
\end{eqnarray}
Some examples of tree graphs and the corresponding spin models are shown in Fig.\ \ref{fig:tree_and_model}.
For instance, the perfect binary tree (PBT) with $n$ layers corresponds to the spin model on an $n$-dimensional hypercube.

We also introduce the block Hamiltonian $\mathcal{H}_v$ for the spin block assigned to a vertex $v$ for later use.
$\mathcal{H}_v$ consists of all the exchange interactions between the spins within the spin block,
\begin{eqnarray}
\mathcal{H}_v &=& \sum_{v' \in B(v)} 2 J_{v'} \tilde{\bm S}_{c_{\rm L}(v')} \cdot \tilde{\bm S}_{c_{\rm R}(v')}
\nonumber \\
&=& \sum_{v' \in B(v)} 2 J_{v'} \sum_{i \in c_{\rm L}(v')}\sum_{j \in c_{\rm R}(v')}
{\bf s}_i \cdot {\bf s}_j,
\label{eq:block_Ham}
\end{eqnarray}
where $B(v)$ represents the set of internal vertices in the spin block assigned to the vertex $v$.
Note that the block Hamiltonian for the root vertex is equivalent to the model Hamiltonian $\mathcal{H}$ [Eq.\ (\ref{eq:Ham_totalspin})].

\section{Exact Ground State}\label{sec:ExactGS}
\subsection{Perfect-Binary Tree}\label{subsec:PBT}

In this section, we explain the derivation of the model that has the exact ground state, using the PBT as a typical example.

\begin{figure}
\includegraphics[width=60mm]{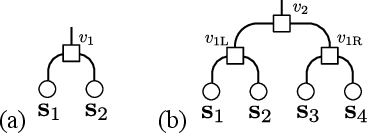}
\caption{
(a) A part of PBT at the first layer from the boundary.
(b) A part of PBT in the two layers from the boundary.
%(c) A part of PBT in two middle layers of PBT.
}
\label{fig:construction}
\end{figure}

First, let us focus on the boundary of the PBT.
Suppose that two $S=1/2$ spins, ${\bm s}_1$ and ${\bm s}_2$, have the same parent vertex $v_1$ and are directly connected by the exchange interaction of $2 J_{v_1}$, as shown in Fig.\ \ref{fig:construction}(a).
The block Hamiltonian of the spin block assigned to the vertex $v_1$, that is nothing but the exchange interaction $2 J_{v_1} {\bm s}_1 \cdot {\bm s}_2$, has the singlet state of those spins as the ground state.
%In the following, we refer to such a condition that every $S=1/2$ spin is directly connected to any one of the other spins by an exchange interaction at the boundary layer of a tree graph, as the singlet-pair-forming condition.

Next, consider the second layer from the boundary of PBT.
There, vertices $v_{1{\rm L}}$ and $v_{1{\rm R}}$, whose spin blocks are composed of two spins $\{ {\bm s}_1, {\bm s}_2 \}$ and $\{ {\bm s}_3, {\bm s}_4 \}$ respectively, are connected by the vertex $v_2$, as shown in Fig.\ \ref{fig:construction}(b).
The Hamiltonian for the spin block of this vertex $v_2$ is written as
\begin{eqnarray}
\mathcal{H}_{v_2} 
&=& 2 J_{v_2} ({\bm s}_1 + {\bm s}_2) \cdot ({\bm s}_3 + {\bm s}_4)
+ 2 J_{v_{1{\rm L}}} {\bm s}_1 \cdot {\bm s}_2
+ 2 J_{v_{1{\rm R}}} {\bm s}_3 \cdot {\bm s}_4
\nonumber \\
&=& J_{v_2} (\tilde{\bm S}_{v_2})^2
+ (J_{v_{1{\rm L}}} - J_{v_2}) (\tilde{\bm S}_{v_{1{\rm L}}})^2
\nonumber \\
&&~~~~~~+ (J_{v_{1{\rm R}}} - J_{v_2}) (\tilde{\bm S}_{v_{1{\rm R}}})^2
+ {\rm const.}.
\end{eqnarray}
Note that $\tilde{\bm S}_{v_2}={\bm s}_1 + {\bm s}_2 + {\bm s}_3 + {\bm s}_4$, $\tilde{\bm S}_{v_{1{\rm L}}} = {\bm s}_1 + {\bm s}_2$, and $\tilde{\bm S}_{v_{1{\rm R}}} = {\bm s}_3 + {\bm s}_4$.
%where $2J_{v_{1{\rm L}}}$ and $2J_{v_{1{\rm R}}}$ are the exchange interaction constants allocated to the vertices $v_{1{\rm L}}$ and $v_{1{\rm R}}$, respectively.
Therefore, if $J_{v_{1{\rm L}}} > J_{v_2}$ and $J_{v_{1{\rm R}}} > J_{v_2}$, one can immediately find that this block Hamiltonian $\mathcal{H}_{v_2}$ has the unique ground state of total spin zero, that is the product state of two singlets on the spin pairs $\{ {\bm s}_1, {\bm s}_2 \}$ and $\{ {\bm s}_3, {\bm s}_4 \}$.
Here, it is also noteworthy that a Hamiltonian $\mathcal{H}_{v_2} - J_p (\tilde{\bm S}_{v_2})^2$ with a certain $J_p$ has the same singlet ground state as that of $\mathcal{H}_{v_2}$ as long as $J_{v_2} > J_p$ holds.

Finally, consider a vertex $v$ at the third or higher layer of PBT from the boundary, as shown in Fig.\ \ref{fig:treegraph}(b).
The Hamiltonian of the spin block of $v$ is written as
\begin{eqnarray}
\mathcal{H}_v &=& 2 J_v \tilde{\bm S}_{c_{\rm L}(v)} \cdot \tilde{\bm S}_{c_{\rm R}(v)}
+ \mathcal{H}_{c_{\rm L}(v)} + \mathcal{H}_{c_{\rm R}(v)}
\nonumber \\
&=& J_v (\tilde{\bm S}_v)^2 + \mathcal{H}_{c_{\rm L}(v)} - J_v (\tilde{\bm S}_{c_{\rm L}(v)})^2
\nonumber \\
&&~~~~~~+ \mathcal{H}_{c_{\rm R}(v)} - J_v (\tilde{\bm S}_{c_{\rm R}(v)})^2,
\label{eq:blockH_v}
\end{eqnarray}
where $\mathcal{H}_{c_{\rm L}(v)}$ and $\mathcal{H}_{c_{\rm R}(v)}$ are the block Hamiltonians of the child vertices of $v$.
Here, suppose that the Hamiltonians $\mathcal{H}_{c_\mu(v)} - J_v (\tilde{\bm S}_{c_\mu(v)})^2$ ($\mu={\rm L, R}$) have a unique ground state with the block spin $\tilde{S}_{c_\mu(v)}=0$.
Then, the block Hamiltonian $\mathcal{H}_v$ [Eq.\ (\ref{eq:blockH_v})] has the unique ground state with its block spin $\tilde{S}_v = 0$, which is the direct product state of the singlet ground states of the blocks of two child vertices.

Iterating the above construction of the block Hamiltonian up to the root vertex leads us to the following conclusion;
if the model (\ref{eq:Ham_totalspin}) corresponding to a PBT satisfies the condition that the exchange interaction constant of every internal vertex $v$ is smaller than the constants of its child vertices $c_{\rm L}(v)$ and $c_{\rm R}(v)$, i.e., $J_v < J_{c_{\rm L}(v)}$ and $J_v < J_{c_{\rm R}(v)}$, the model possesses the unique ground state given by
\begin{eqnarray}
|\Psi \rangle = \prod_{k=1}^{N/2} \otimes | S_{i_1(k) i_2(k)} \rangle,
\label{eq:singlet_productGS}
\end{eqnarray}
where $| S_{i_1(k) i_2(k)} \rangle$ is the spin-singlet dimer state composed of two spins ${\bm s}_{i_1(k)}$ and ${\bm s}_{i_2(k)}$.
For the case of PBT, $\{ {\bm s}_{i_1(k)}, {\bm s}_{i_2(k)}\}$ is a pair of spins connected via a vertex at the boundary layer.

\subsection{Generic Tree Graph}\label{subsec:genericTree}

Next, we extend the derivation of the exact ground state for the model of PBT discussed in the previous section to the cases of generic tree graphs.
To that end, we rewrite the exchange interaction $h_v$ [Eq.\ (\ref{eq:int_vertex})] assigned to each internal vertex $v$ as
\begin{eqnarray}
h_v = J_v \left[ (\tilde{\bm S}_v)^2 - (\tilde{\bm S}_{c_{\rm L}(v)})^2 - (\tilde{\bm S}_{c_{\rm R}(v)})^2 \right].
\end{eqnarray}
Using this expression and considering the connection of a tree graph, one can express the model Hamiltonian corresponding to a generic binary tree graph as
\begin{eqnarray}
\mathcal{H} = \sum_v \left[ J_v - J_{p(v)} \right] (\tilde{\bm S}_v)^2 + {\rm const.}.
\label{eq:Ham_v-pv}
\end{eqnarray}
Here, $J_{p(v)}$ is a half of the exchange interaction constant allocated to the parent vertex of $v$, and $J_{p(v)}=0$ for the root vertex.
From the expression of Eq.\ (\ref{eq:Ham_v-pv}), it immediately follows that if the condition that the exchange constant at a vertex $v$ is larger than that of its parent vertex, $J_v > J_{p(v)}$, holds for every $v$, the magnitude of the block spin $\tilde{S}_v$ in the ground state takes the smallest value, that is $\tilde{S}_v=0$ ($1/2$) for the block with an even (odd) number of spins.

\begin{figure}
\includegraphics[width=60mm]{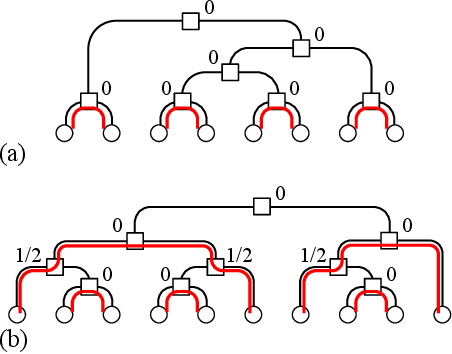}
\caption{
(Color online) 
Correspondence between tree graphs and the ground states.
(a) Tree graph containing only vertices whose spin block consists of an even number of spins.
(b) Tree graph containing vertices whose spin block consists of an odd number of spins.
The numbers indicated for vertices denote the magnitude of the block spin $\tilde{S}_v$ for each vertex in the ground state.
Red lines represent the spin pairs forming the singlet state in the ground state.
}
\label{fig:Tree_GS}
\end{figure}

Now, we discuss the ground state of the model (\ref{eq:Ham_v-pv}) for generic tree graphs.
During this discussion, we assume that the condition $J_v > J_{p(v)}$ is satisfied for all internal vertices $v$.
First, let us consider the tree graph in which every $S=1/2$ spin is directly connected to any one of the other spins via a single vertex at the boundary layer.
See Fig.\ \ref{fig:Tree_GS}(a) for an example.
For such a tree graph, the spin block for every internal vertex $v$ contains an even number of spins.
Therefore, the ground state of the corresponding model has $\tilde{S}_v=0$ for all the internal vertices.
This ground state is nothing but the product state of singlet dimers, Eq.\ (\ref{eq:singlet_productGS}).

\begin{figure}
\includegraphics[width=60mm]{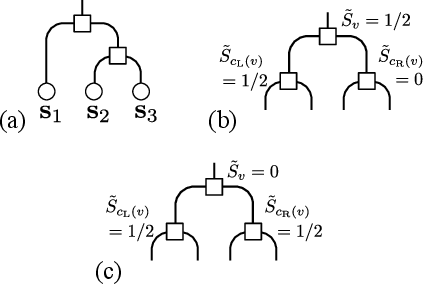}
\caption{
(a) A part of a tree graph representing a three-spin block consisting of ${\bm s}_1$, ${\bm s}_2$, and ${\bm s}_3$.
(b) and (c) represent a part of a tree graph where two spin blocks of vertices $c_{\rm L}(v)$ and $c_{\rm R}(v)$ are merged at the vertex $v$.
$\tilde{S}_v$, $\tilde{S}_{c_{\rm L}(v)}$, and $\tilde{S}_{c_{\rm R}(v)}$ denote the magnitude of the block spins in the ground state.
}
\label{fig:odd_spin_block}
\end{figure}

Next, consider the tree graphs containing internal vertices whose spin block is composed of an odd number of spins.
Here, let us closely look at the ground state of such a spin block with an odd number of spins.
The smallest such block is the three-spin block shown in Fig.\ \ref{fig:odd_spin_block}(a).
The ground state of its block Hamiltonian is the product state of the $S=1/2$ doublet of the bare spin ${\bm s}_1$ and the singlet state formed by the spins ${\bm s}_2$ and ${\bm s}_3$.
It should be noted that the $S=1/2$ degree of freedom of the spin block stems solely from the bare spin ${\bm s}_1$.
Then, consider the case that two spin blocks are merged at a vertex $v$, where one block contains an odd number of spins and has a single $\tilde{S}_{c_{\rm L}(v)}=1/2$ doublet as the unique ground-state doublet, while the other is the block of an even number of spins having the unique singlet ground state with $\tilde{S}_{c_{\rm R}(v)}=0$ [see Fig.\ \ref{fig:odd_spin_block}(b)].
The ground state of such a merged block is a single $\tilde{S}_v=1/2$ doublet, which is the product of the ground states of the individual blocks.
Therefore, the ground state of a spin block containing an odd number of spins is a single $\tilde{S}_v=1/2$ doublet, which can be traced back to a single bare spin in the block.
Such a ground-state doublet survives even when the block is merged with a block with an even number of spins.
Then, when the spin block is merged with the other block with an odd number of spins [Fig.\ \ref{fig:odd_spin_block}(c)], the merged spin block has the singlet ground state in which two bare spins representing the ground-state doublet of each block form a singlet pair.
We thereby find that the model corresponding to the tree graph considered here has a product state of singlet dimers Eq.\ (\ref{eq:singlet_productGS}) as its unique ground state.
The pattern of the singlet-dimer distribution depends on the tree graph; Fig.\ \ref{fig:Tree_GS} presents examples.

From the above argument, the conclusion follows: 
For an arbitrary binary tree graph, the corresponding spin model Eq.\ (\ref{eq:Ham_totalspin}) possesses a singlet-dimer product state Eq.\ (\ref{eq:singlet_productGS}) as the unique ground state, if the condition $J_{p(v)}<J_v$ holds for all internal vertices, in other words, if the interaction strength $J_v$ increases from the root vertex to the boundary of the tree graph.
The spatial pattern of the singlet dimers can be determined diagrammatically using the tree graph, as depicted in Fig.\ \ref{fig:Tree_GS}.

\subsection{Degenerate Ground States}\label{subsec:degenerateGS}

Consider the case where the condition that the exchange interaction constant at a vertex is stronger than the one at its parent vertex, $J_{p(v)} < J_v$, is relaxed to $J_{p(v)}=J_v$ at some vertices.
In this case, in the model Hamiltonian written as Eq.\ (\ref{eq:Ham_v-pv}), the terms proportional to $(\tilde{\bm S}_v)^2$ for those vertices vanish, and the magnitude of the block spin $\tilde{S}_v$ has no effect on the energy of the model.

As a typical example, consider a PBT where $J_{p(v)}=J_v$ holds at the vertices in the first layer from the boundary of PBT, while $J_{p(v)} < J_v$ is satisfied for the vertices in the second and higher layers.
In the ground state of this model, the magnitudes of all the block spins assigned to the vertices in the second and higher layers are zero, yielding the ground state of the product of the singlet states for the four-spin blocks assigned to the vertices in the second layer,
\begin{eqnarray}
|\Psi \rangle = \prod_{k=1}^{N/4} \otimes |S_{i_1(k) i_2(k) i_3(k) i_4(k)} \rangle,
\end{eqnarray}
where $|S_{i_1(k) i_2(k) i_3(k) i_4(k)} \rangle$ is one of the singlet states composed of the spins in the four-spin blocks.
(Here, we suppose that $N$ is a multiple of four.)
Since the four-spin block has two singlet states, the ground states of the entire model are $2^{N/4}$-fold degenerate.

Similarly, for the three-spin block shown in Fig.\ \ref{fig:odd_spin_block}(a), if we set $J_{p(v)}=J_v$ at the vertex $v$ connecting ${\bm s}_2$ and ${\bm s}_3$, the three spins are coupled by the exchange interactions of the same strength.
Such a triangle unit has the four-fold degenerate ground states composed of two $S=1/2$ doublets characterized by the spin and chirality degrees of freedom.
One can adopt such three-spin units as building blocks to construct the degenerate ground states.

In the above manner, for arbitrary tree graphs, one can construct a model with massively degenerate ground states by setting $J_{p(v)}=J_v$ at the lower layer(s).

\section{Summary}\label{sec:summary}

In summary, we have introduced a protocol for constructing an $S=1/2$ antiferromagnetic Heisenberg model with the exact ground state based on a tree graph.
The protocol determines the spatial configuration of exchange interactions of the model, which has a one-to-one correspondence with the tree graph.
Then, we have proven that when the exchange interactions allocated to vertices satisfy the condition that their strength increases as one moves from the root vertex to the boundary of the tree graph, the model possesses a product state of singlet dimers as its unique ground state.
It has also been shown that one can construct a model possessing the massively degenerate ground states by relaxing the condition of the strength of the interactions.

Although we have focused on binary tree graphs and $S=1/2$ spin models, the protocol we proposed can be extended to $m$-ary tree graphs and the models with an arbitrary spin size $S$ straightforwardly.

The results of this study introduce a class of quantum spin models with the exact ground state by employing the correspondence with tree graphs.
We hope the results will stimulate further studies, such as the classification of quantum spin models using graph theory.
Realization of the models, which allow us to arrange singlet dimers as desired in their ground states spatially, would be an interesting future research.

%\begin{acknowledgment}
\acknowledgments
We thank Kouichi Okunishi and Tomotoshi Nishino for fruitful discussions.
T.H. was supported by JSPS KAKENHI Grant Number JP24K06881.

%\end{acknowledgment}

%\appendix
%\section{}

$^*$hikihara@gunma-u.ac.jp

%\bibliography{ExactGS-Tree_Ref}

\end{document}